\documentclass[12pt,preprint]{aastex}

\newcommand       \AU           {\,{\rm AU}}

\newcommand	  \g		{\,{\rm g}}
\newcommand       \K            {\,{\rm K}}

\newcommand	  \s		{\,{\rm s}}

\newcommand	  \yr		{\,{\rm yr}}

\newcommand       \simlt        {\lesssim}

\newcommand       \mum          {\,{\rm \mu m}}
\newcommand	  \Teff	        {T_{\rm eff}}
\newcommand	  \amin	        {a_{\rm min}}
\newcommand	  \amax	        {a_{\rm max}}
\newcommand	  \rmin	        {r_{\rm min}}
\newcommand	  \rmax	        {r_{\rm max}}

\newcommand	  \rp           {r_{\rm p}}

\newcommand       \Lsun         {L_\odot}

\newcommand       \Lstar        {L_\star}
\newcommand       \simali       {\sim\,}
\newcommand{\figwidth}{6.0in}

\shorttitle{Mid-IR AO Imaging of CH Cygni}
\shortauthors{Biller et al.}

\begin{document}


\title{Resolving the Dusty Circumstellar Structure of the Enigmatic Symbiotic
Star CH Cygni with the MMT Adaptive Optics System{\footnotemark}}

\footnotetext{The results presented here made use of 
              the of MMT Observatory, a facility jointly 
              operated by the University of Arizona 
              and the Smithsonian Institution.}


\author{Beth A. Biller$^1$, Laird M. Close$^1$, Aigen Li$^{2}$, 
Massimo Marengo$^3$, John H. Bieging$^1$, Phil M. Hinz$^1$, 
William F. Hoffmann$^1$, Guido Brusa$^1$, and Doug Miller$^1$}

%\email{bbiller@as.arizona.edu}

\affil{$^1$Steward Observatory, University of Arizona, Tucson, AZ 85721;
       {\sf bbiller@as.arizona.edu}}
\affil{$^2$Department of Physics and Astronomy,
           University of Missouri, Columbia, MO 65211}
\affil{$^3$Harvard-Smithsonian Center for Astrophysics, 
       60 Garden Street, Cambridge, MA 02138}



\begin{abstract} 
We imaged the symbiotic star CH Cyg and two PSF calibration stars
using the unique 6.5\,m MMT deformable secondary adaptive optics
system. Our high-resolution (FWHM\,=\,0.3$\arcsec$), very high
Strehl ($98\pm2\%$) mid-infrared (9.8 and 11.7$\mum$) images 
of CH Cyg allow us to probe finer length scales than ever before 
for this object.  CH Cyg is significantly extended compared to 
our unresolved PSF calibration stars ($\mu$ UMa and $\alpha$ Her) 
at 9.8 and 11.7$\mum$.  We estimated the size of the extension by 
convolving a number of simple Gaussian models 
with the $\mu$ UMa PSF and determining which model 
provided the best fit to the data.
Adopting the Hipparcos distance for this object of 270\,pc, 
we found a nearly Gaussian extension with a FWHM at 9.8$\mum$ 
of $\simali$40.5$\pm$2.7$\AU$ (0.15$\pm$0.01$\arcsec$) 
and a FWHM at 11.7$\mum$ of 
45.9$\pm$2.7$\AU$ (0.17$\pm$0.01$\arcsec$).
After subtracting out the Gaussian component of the emission
(convolved with our PSF), 
we found a faint $\simali$0.7$\arcsec$ asymmetric extension 
which peaks in flux $\simali$0.5$\arcsec$ north of the stars.  
This extension is roughly coincident with the northern
knotlike feature 
seen in HST WFPC2 images obtained in 1999.
\end{abstract}


\section{Introduction}
Symbiotic stars are a class of interacting binaries consisting of a hot 
component (usually a white dwarf) which ionizes the stellar wind of 
a cool component (generally a red giant or Mira, see Kenyon 2000).  
With typical component separations of $\simali$1--50$\AU$, 
these complex systems are responsible for some of the most 
dramatic Galactic jet sources \citep{eyr02}.  

CH Cyg is one of the most enigmatic symbiotic stars -- 
previous to 1963, CH Cyg showed no variable behavior 
and was actually used as a M6III spectral type calibrator
(see Kenyon 2001, for background details~on CH Cyg's properties).
Since 1963, CH Cyg has gone through multiple outbursting phases, 
producing bipolar radio jets in 1984
\citep{tay86}, 1992 \citep{kcm98}, and 1998 \citep{cro01}, 
as well as X-ray jets \citep{gs04}, and large ionized outflows 
\citep[$\sim$10$\arcsec$~in size,~see][]{eyr02,cor01}.
B band flickering studies also point to the existence of an unstable
accretion disk around the hot white dwarf 
component \citep[][]{sk031,sk032}.

The CH Cyg system contains at least one red giant/white dwarf pair 
\citep{mik87}.  Currently, there is ongoing debate \citep{sk032} as to 
whether the CH Cyg system is binary
or the only known symbiotic triple star (consisting of either 
a red giant/white dwarf pair with a G star companion \citep{hin93} 
or a red giant/white dwarf pair with an additional
red giant companion \citep{sko96}).
For the purposes of this paper, we adopt the interpretation of the CH Cyg
system as an inner red giant/white dwarf pair with 
a period of 756 days \citep{hin93} and an outer red giant 
with a period of 14.5 years \citep{sko96}. 
All three stars are viewed nearly edge on and the outer giant 
periodically eclipses the inner binary.

As an unresolved triple system, CH Cyg is very difficult to study. 
In order to completely understand the system it is necessary to 
disentangle the contribution of each component to the behavior 
of the system as a whole.
Most studies of CH Cyg focus on the contribution 
of the hot white dwarf component to the system -- jets \citep{gs04}, 
ionized outflows \citep[][]{eyr02,cor01}, and hot component disk flickering 
\citep[][]{sk031,sk032} primarily trace the properties of the hot component 
and the matter accreting on to it.  
In fact, the hot component dominates the total source properties 
at many wavelengths -- to study the cool components, 
it is important to find a wavelength 
at which the cool components dominate over the hot component.  
To this end, we chose to study CH Cyg at mid-infrared (mid-IR) wavelengths, 
where the contribution of the hot component is negligible 
($<$\,5$\%$ of the total emission, even during active periods). 
While near-IR and mid-IR photometry has previously 
been undertaken for CH Cyg 
(Taranova \& Shenavrin 2000, 2004;
Bogdanov \& Taranova 2001),
%[\citet{ts00},\citet{ts04},\citet{bt01}],
CH Cyg has not previously been resolved at these wavelengths, 
since sufficient spatial resolution at mid-IR wavelengths has 
not been available until recently 
(Close et al.\ 2003, Biller et al.\ 2005).
Using the novel technique of adaptive optics at mid-IR wavelengths,
we can directly image the emission from 
the cool component of the system by itself 
for the first time and study any circumstellar material 
on finer spatial scales than before.

\section{Observations and Data Reduction}
Data were taken on the night of 2003 May 13 (UT) 
at the 6.5\,m MMT using the adaptive secondary 
mirror AO system \citep{wil03} with the BLINC-MIRAC3 
camera (Hoffman et al.\ 1998; Hinz et al.\ 2000).  
Currently all other existing AO systems use 
a warm reimaged adaptive mirror and require 
several additional warm optical
surfaces between the secondary and the detector.
With the unique MMT adaptive secondary, 
we remove many of these warm optical surfaces from the optical
path compared to a standard AO system.  
This results in a considerably reduced 
thermal background at the mid-IR, 
thus making adaptive optics
possible for mid-IR observations.  
The adaptive secondary corrected
the first 52 system modes and achieved Strehl ratios 
as high as 0.97$\pm$0.03 from 8.8 to 18$\mum$. 
These Strehl ratios are the highest ever presented in 
the literature for a large ground based telescope 
(Close et al.\ 2003, Biller et al.\ 2005).
For details on the MMT Adaptive Secondary AO system, 
see \citet{wil03}.
We utilized the 128$\times$128 SiAs BIB 2--20$\mum$ 
MIRAC3 camera \citep{hof98}. 
The $0.088\arcsec$/pixel scale was used with the
9.8 and 11.7$\mum$ 10\% bandwidth filters. 
To remove thermal and detector instabilities 
we chopped at 1\,Hz with an internal pupil-plane 
cold chopper in the interface dewar BLINC 
\citep{hin00} between the AO system and MIRAC3.

\begin{figure}
\includegraphics[angle=0,width=\columnwidth]{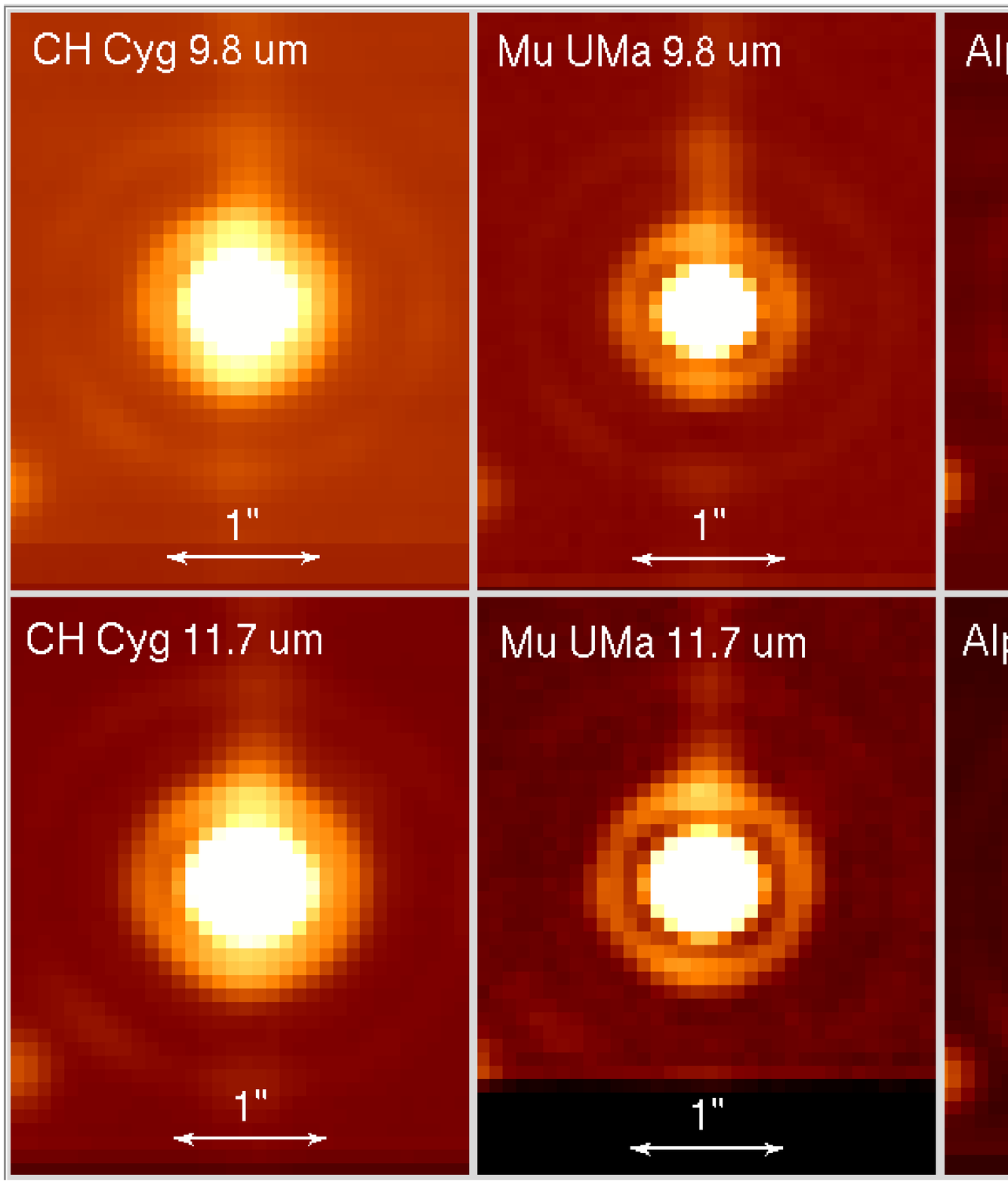}
\caption{
         The 9.8 and 11.7$\mum$ images of CH Cyg and PSF stars 
         $\mu$ UMa, $\alpha$ Her, and AC Her as observed at the MMT.  
         The first, second, and third Airy rings are visible
         in these images. CH Cyg appears slightly extended 
         compared to the PSF stars -- while the PSF stars show 
         a peak distinct from the first Airy ring, 
         the slightly extended emission from CH Cyg blurs into 
         the first Airy ring. The faint point source in the far 
         lower left of each MMT image is a MIRAC3 ghost.}
\label{fig:images}
\end{figure}

We observed with the AO system locked continuously on CH Cyg.  
The $15^{\circ}$ tilted BLINC dewar window 
is a high quality dichroic which reflected 
the visible light ($\lambda$\,$<$\,1$\mum$) 
to the AO wavefront sensor and transmitted 
the IR through BLINC to MIRAC3. 
Since the internal chopper in BLINC was past the dichroic, 
continuous 1\,Hz chopping did not affect the visible light 
beam and hence the AO lock was unaffected. To further 
calibrate the background (in addition to chopping) 
we nodded $\simali$6--8$\arcsec$ in the telescope's 
azimuth direction (the horizontal direction in 
Figure \ref{fig:images}) every minute. 
The internal chopper was set to run in the altitude direction
(the vertical direction) with a chop throw of $\simali$20$\arcsec$.
The derotator was disabled during these observations 
to help minimize the residual background structure.

The $0.505\arcsec$ PA\,=\,269$^{\circ}$ 
Washington Double Star catalog astrometric binary 
WDS 02589+2137 BU was observed during a later
run (2003, November 25 UT) and used to calibrate 
the camera's orientation and its $0.088\arcsec$/pixel platescale.

For the 9.8 and 11.7$\mum$ filters 
we obtained 2$\times$1 minute coadded 
chop differenced images 
(one image from each nod) for CH Cyg.  
We utilized a custom IRAF script to reduce this mid-IR data
\citep{bil03}. The script produced two background subtracted images
by subtracting the image taken at the first nod position with that 
taken at the second, and vice versa.  
Images were then bad pixel corrected, flat-fielded, 
cross-correlated, and aligned to a reference nod image
(to an accuracy of $\sim$0.02 pixels). Mid-IR images of the 
PSF calibration stars ($\mu$ UMa and $\alpha$ Her, 
both of these stars are binary but appear unresolved to MIRAC, 
see Close et al.\ [2003] for a discussion of PSF suitability) 
were obtained and reduced in an identical manner to CH Cyg.   
Reduced images of CH Cyg and the two PSF stars 
are presented in Figure\,\ref{fig:images}.
In order to directly compare our science images 
with the PSF images, we have not rotated the images 
to place north up -- azimuth is vertical and altitude 
is horizontal.  At the time of the observations, 
the Parallactic Angle (PA) was roughly 160$^{\circ}$.

\begin{figure}
\includegraphics[width=5in]{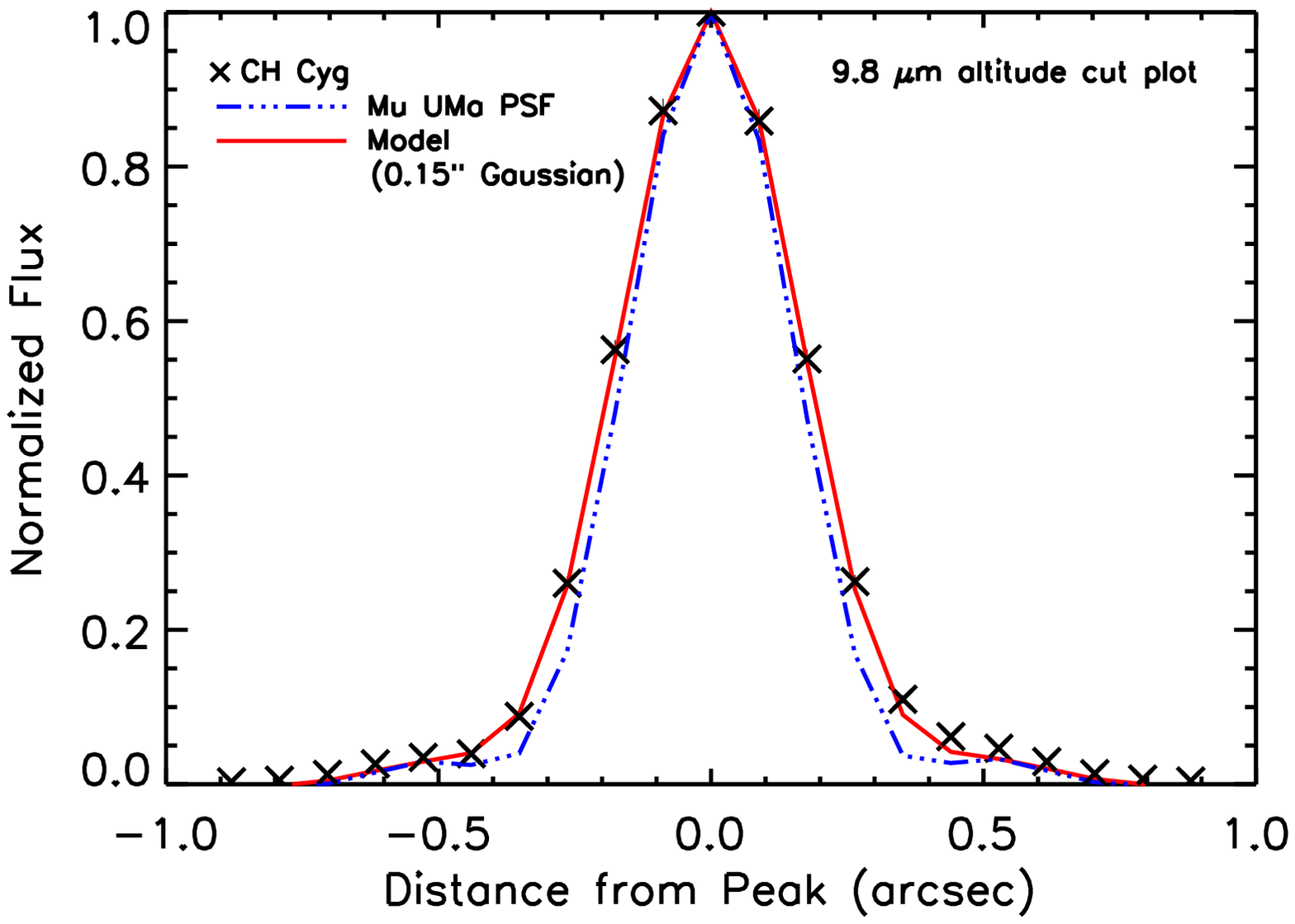} \\
\includegraphics[width=5in]{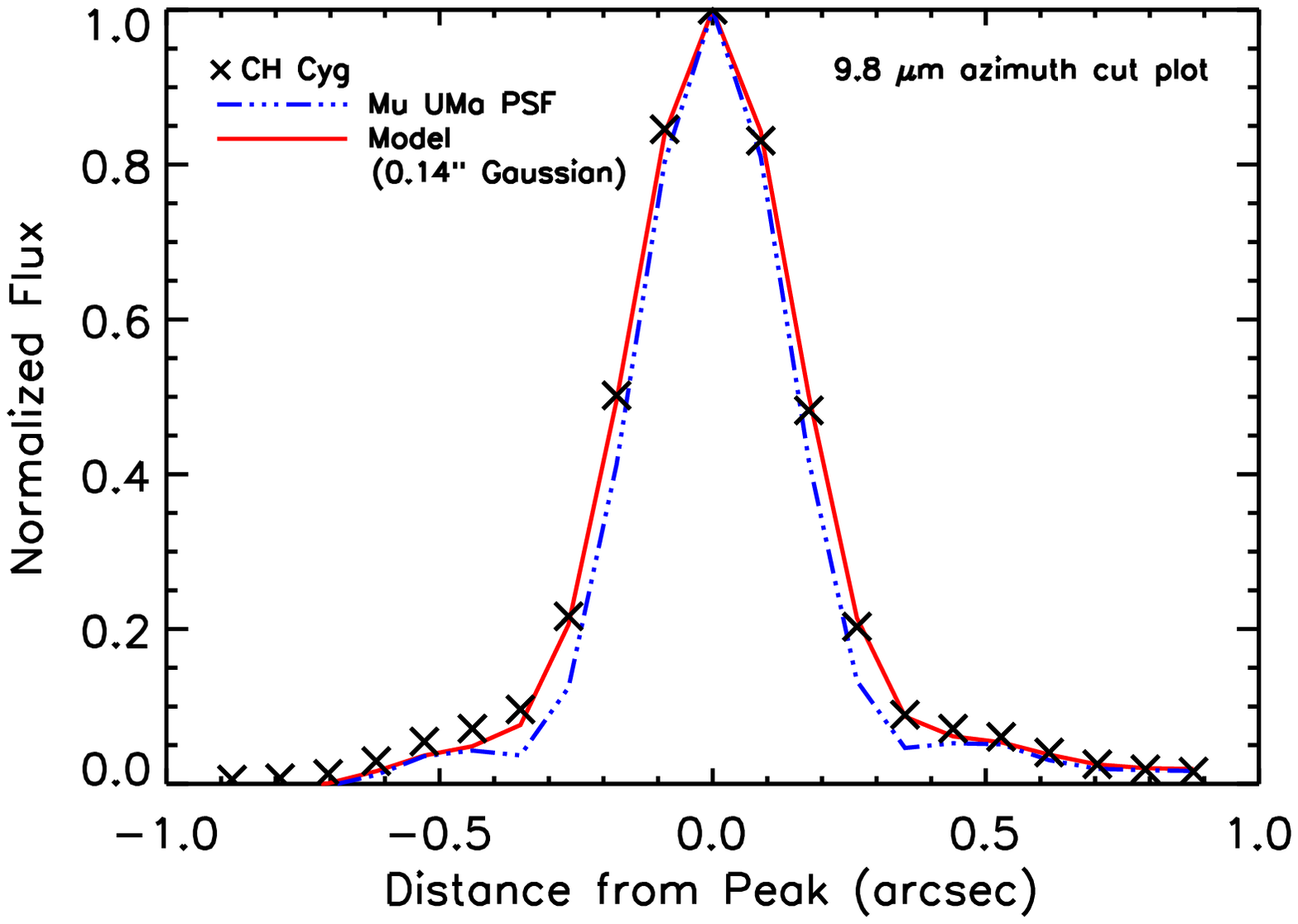} \\
\caption{One dimensional cut plots along the altitude (top) 
         and azimuth (bottom) directions for the 9.8$\mum$ images 
         of CH Cyg and the PSF star $\mu$ UMa.  The CH Cyg data 
         is plotted as crosses, the $\mu$ UMa PSF is plotted as 
         a dotted blue line, and the best fit Gaussian model 
         convolved with the $\mu$ UMa PSF is plotted as a solid 
         red line.  Error bars are small compared to the size of 
         the plot symbols.  CH Cyg appears slightly but notably 
         extended relative to the $\mu$ UMa PSF.}
\label{fig:cutplots9.8}
\end{figure}

\begin{figure}
\includegraphics[width=5in]{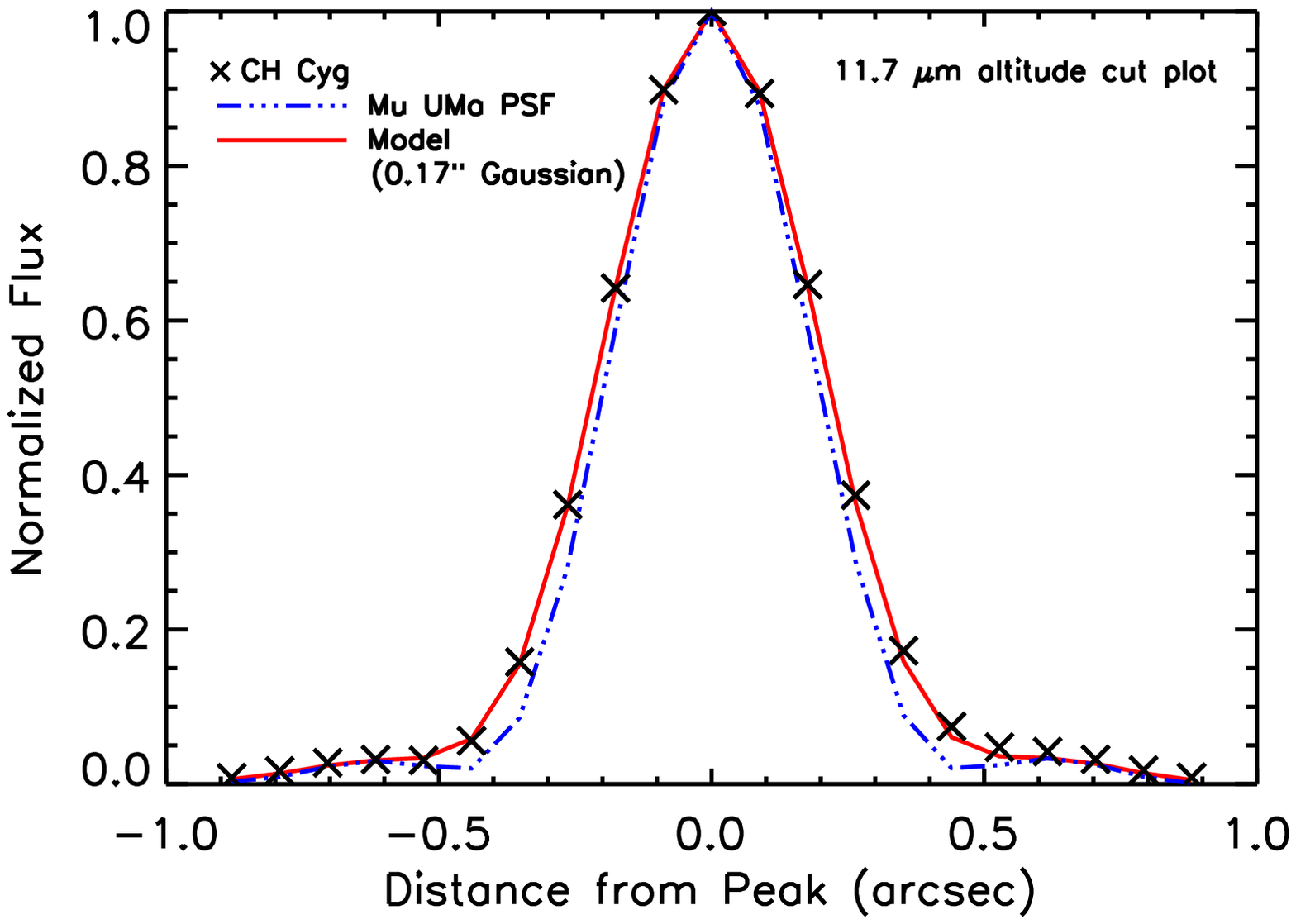} \\
\includegraphics[width=5in]{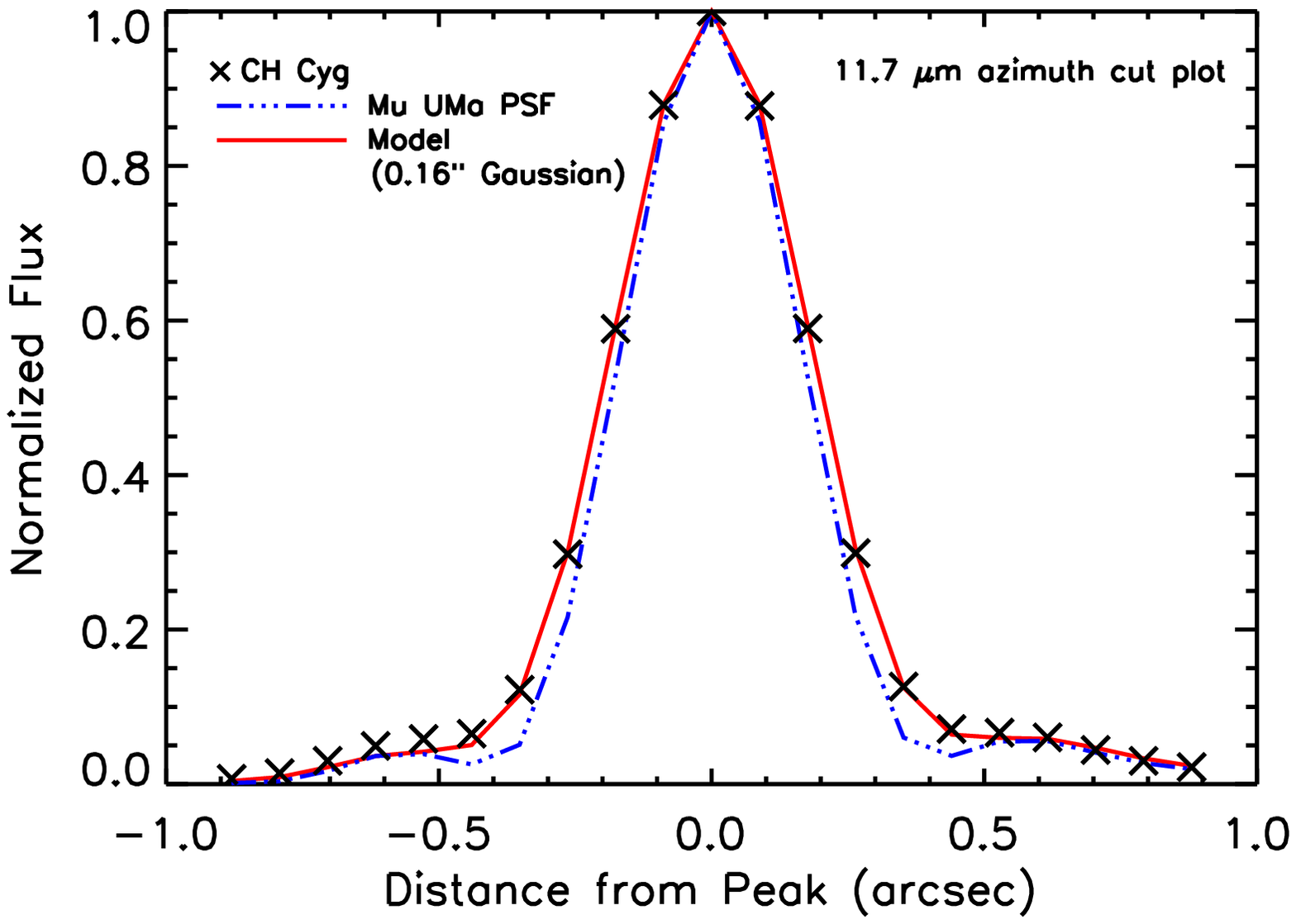} \\
\caption{One dimensional cut plots along the altitude (top) 
         and azimuth (bottom) directions for the 11.7$\mum$ images 
         of CH Cyg and PSF star $\mu$ UMa.  The CH Cyg data 
         is plotted as crosses, the $\mu$ UMa PSF is plotted 
         as a dotted blue line, and the best fit Gaussian model 
         convolved with the $\mu$ UMa PSF is plotted as a solid 
         red line.  Error bars are small compared to the size of 
         the plot symbols.  CH Cyg appears slightly but notably 
         extended relative to the $\mu$ UMa PSF.}
\label{fig:cutplots11.7}
\end{figure}

\begin{figure}
\includegraphics[angle=0,width=5in]{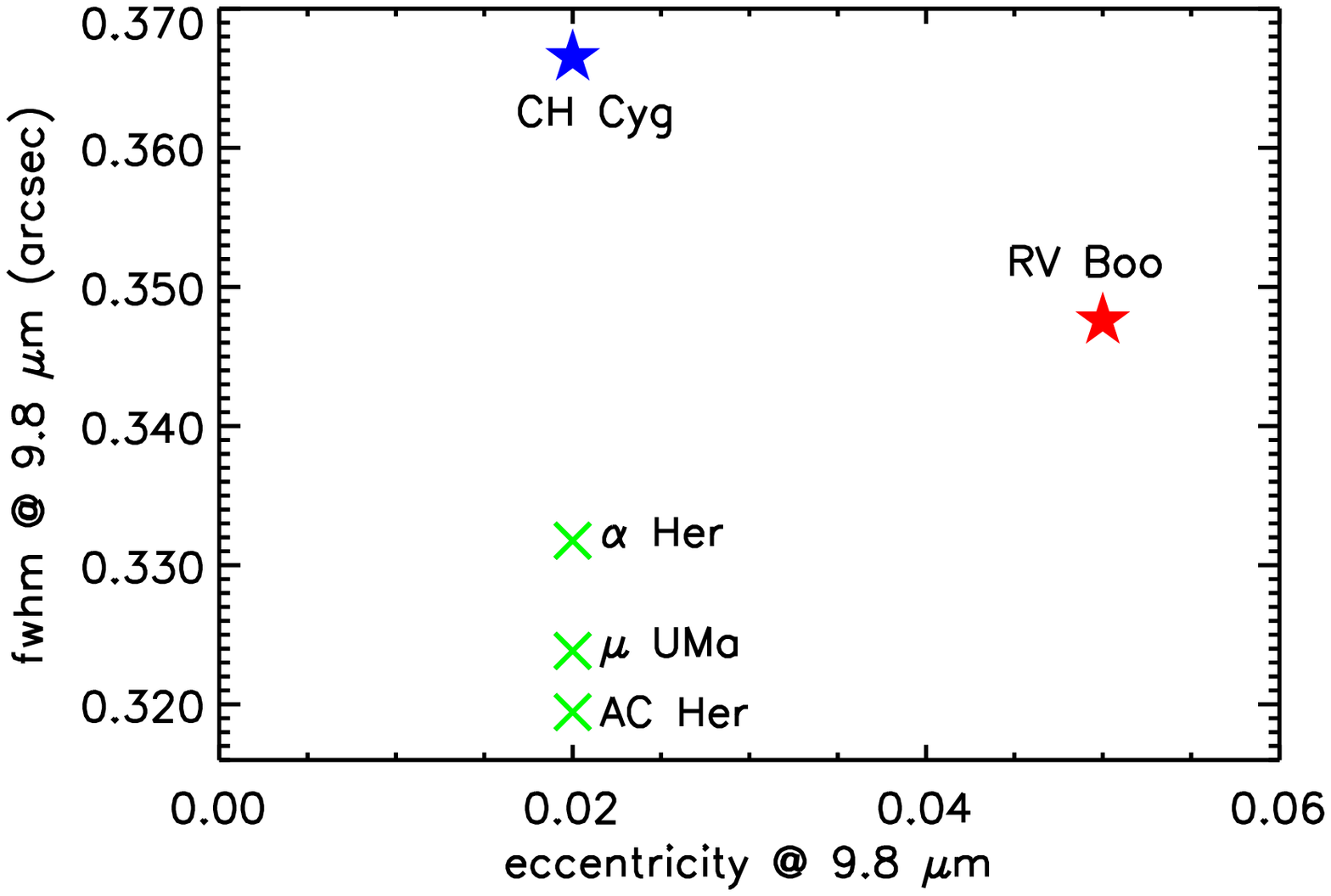}
\includegraphics[angle=0,width=5in]{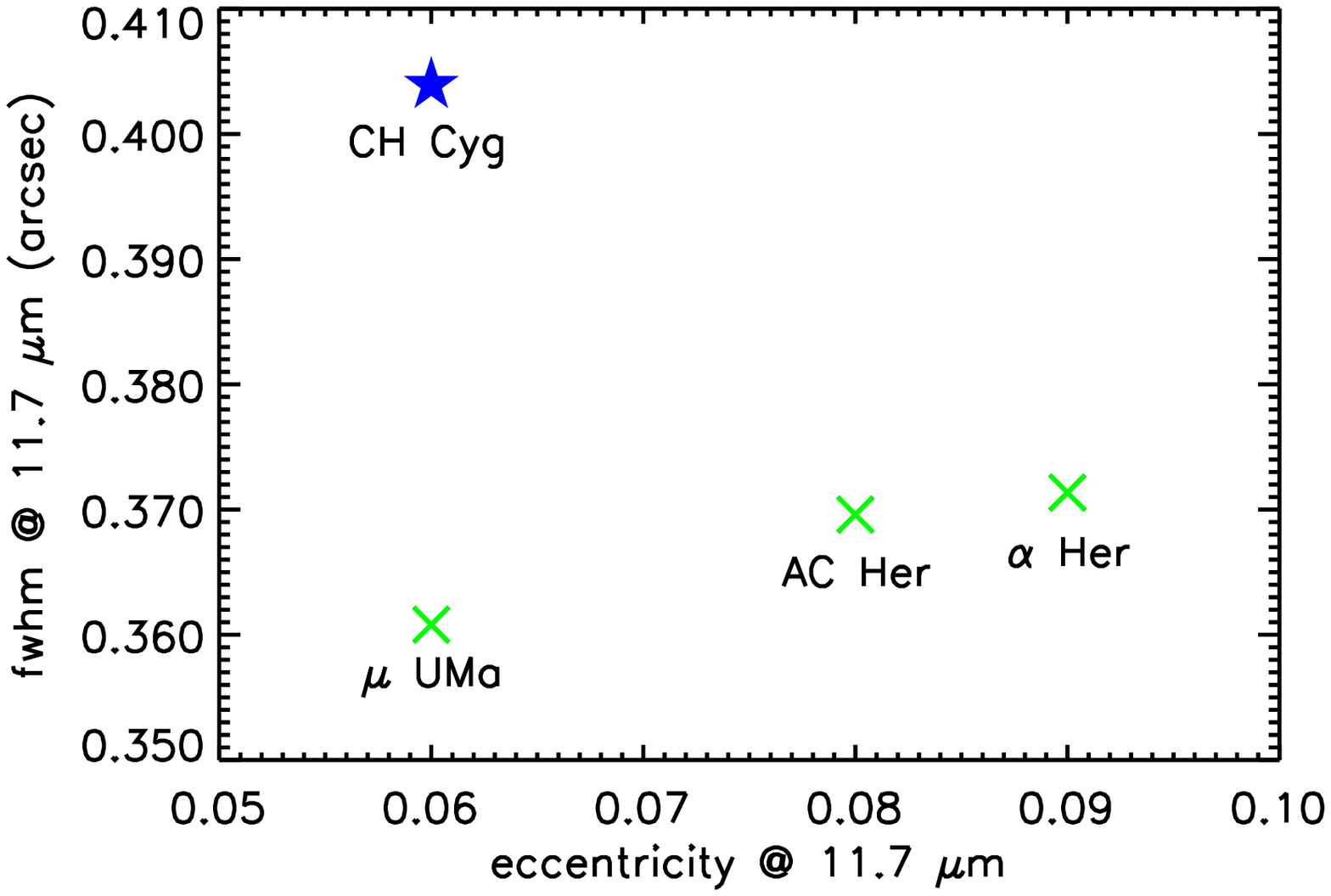}
\caption{The 9.8$\mum$ (top) and 11.7$\mum$ (bottom) FWHM 
         and ellipticity of CH Cyg and the PSF stars $\mu$ UMa 
         and $\alpha$ Her. CH Cyg appears slightly extended 
         in FWHM compared to the PSF stars.} 
\label{fig:FWHM}
\end{figure}

The CH Cyg images appear slightly yet significantly 
extended compared to images of the two PSF stars.  
To highlight this extension, we present one-dimensional 
normalized cut plots in both altitude and azimuth directions 
across the peak for the 9.8 and 11.7$\mum$ images 
of CH Cyg and PSF star $\mu$ UMa in 
Figures\,\ref{fig:cutplots9.8} and \ref{fig:cutplots11.7}.
Three rows/columns were averaged together
for each cut plot to increase signal to noise.  
Since CH Cyg is a very bright object (565\,Jy at 12$\mum$)
and errors are dominated by photon noise, 
error bars for these plots are considerably 
smaller than the size of the plot symbols 
and have not been plotted.  In all the altitude and 
azimuth cutplots at 9.8 and 11.7$\mum$, 
the CH Cyg profiles are noticeably extended compared to 
those for the $\mu$ UMa PSF.  
We measured the FWHM, ellipticity, and positional angle
of any such ellipticity for all the images in 
Figure\,\ref{fig:images}, as well as for images of RV Boo, 
an AGB star with a small extended dust disk observed 
the same night \citep{bil03}. 
In Figure\,\ref{fig:FWHM} we plot the ellipticity 
and FWHM for our datasets.  
CH Cyg has a similar eccentricity but a significantly 
higher FWHM than the PSF stars.
Thus, CH Cyg looks clearly extended compared 
to the PSF stars. 

In order to detect the gas counterpart to the mid-IR emitting dust, 
CH Cyg was observed in the CO\,(1--0) transition at the Arizona Radio 
Observatory 12\,m telescope on Kitt Peak on the night of 29 June 2005. 
No CO\,(1--0) was detected in 4 hours of integration time.

\section{Analysis}
\begin{figure}
\includegraphics[angle=0,width=\columnwidth]{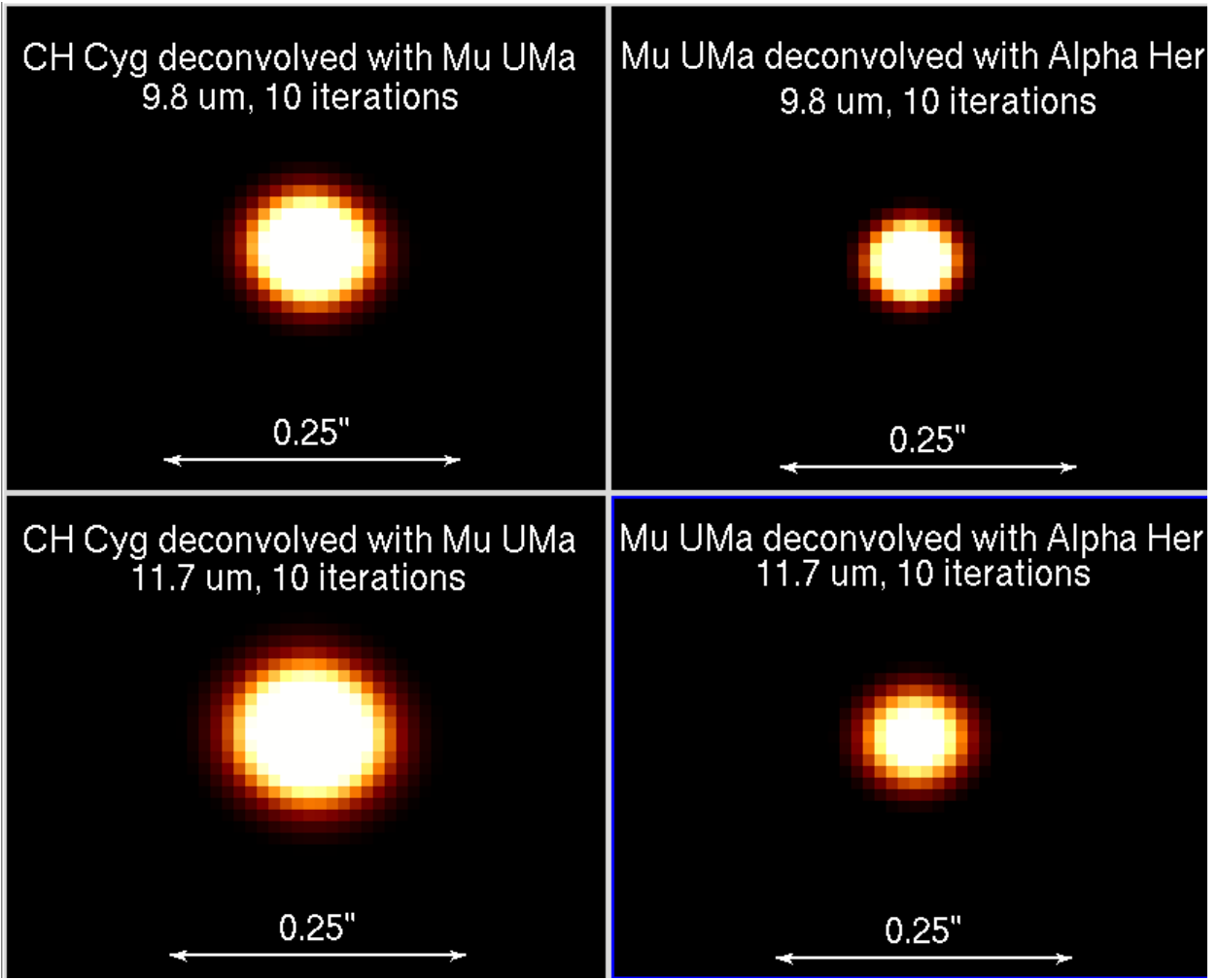}
\caption{Deconvolved images of CH Cyg. The plate scale 
        (after magnification) is 29.3\,mas\,pixel$^{-1}$.  
        Deconvolved images of the $\mu$ UMa PSF 
        (deconvolved using $\alpha$ Her as a PSF) 
        is shown on the right for comparison.  
        CH Cyg appears slightly extended after deconvolution 
        compared to the PSF stars.}
\label{fig:deconvim}
\end{figure}

In order to study small-scale circumstellar structure 
around CH Cyg, we deconvolved our high S/N 
(S/N\,$>$\,1000) images of CH Cyg using 
both the PSF stars $\mu$ UMa and $\alpha$ Her 
(see Biller et al.\ 2005, for details on PSF star selection)
with the IRAF Lucy deconvolution task \citep{lucy74}.  
So as not to introduce noise amplification artifacts 
due to deconvolution, we deconvolved images for only 10 iterations.  
To test whether artifacts were being introduced 
by the Lucy algorithm, we also deconvolved images 
of the $\mu$ UMa PSF with those of the $\alpha$ Her PSF 
in order to establish the expected result of deconvolving
one point source by another point source.  
Deconvolved images of CH Cyg and the PSF stars 
appear in Figure\,\ref{fig:deconvim}.  
Due to the very high Strehl ($\simali$0.98) 
and high signal-to-noise ratio in our PSF images 
we could detect low contrast structure on ``super-resolution'' 
scales of $\simali$0.2$\arcsec$ with the IRAF Lucy
deconvolution task \citep{bil03}.  
After deconvolution,  CH Cyg still appears significantly 
extended compared to the deconvolved PSF images. 

To estimate the size of the extension around CH Cyg, 
we convolved a range of simple Gaussian dust envelope
models (with FWHMs of 0.1, 0.14, 0.15, 
0.16, 0.18, and 0.2$\arcsec$ before convolution) 
with the $\mu$ UMa PSF. We determined the best fit model 
by comparing model cut plots with data cut plots; 
the best fit model minimizes the standard deviation 
of [{\it cutplot}(data)\,-\,{\it cutplot}(model)].  
At 9.8$\mum$, the 0.15$\arcsec$ model provided 
the best fit to the altitude cut plot, 
while the 0.14$\arcsec$ model provided
the best fit to the azimuth cut plot.  
At 11.7$\mum$, the 0.17$\arcsec$ 
model provided the best fit to the altitude cut plot, 
while the 0.16$\arcsec$ model provided the best fit to 
the azimuth cut plot.  Best fit model cut plots are presented 
in Figures\,\ref{fig:cutplots9.8},\ref{fig:cutplots11.7}.  
The slightly broader cut plot profile along the altitude 
direction in both these cases is probably due to 
instrumental effects and is not physical.  
Utilizing the Hipparcos distance of $\simali$270\,pc, 
we adopt a FWHM at 9.8$\mum$ of $\simali$40.5$\pm$2.7\,AU 
(0.15$\pm$0.01$\arcsec$), 
and a FWHM at 11.7$\mum$ of 45.9$\pm$2.7\,AU 
(0.17$\pm$0.01$\arcsec$).     

\begin{figure}
\begin{tabular}{c}
\includegraphics[angle=0,width=\columnwidth]{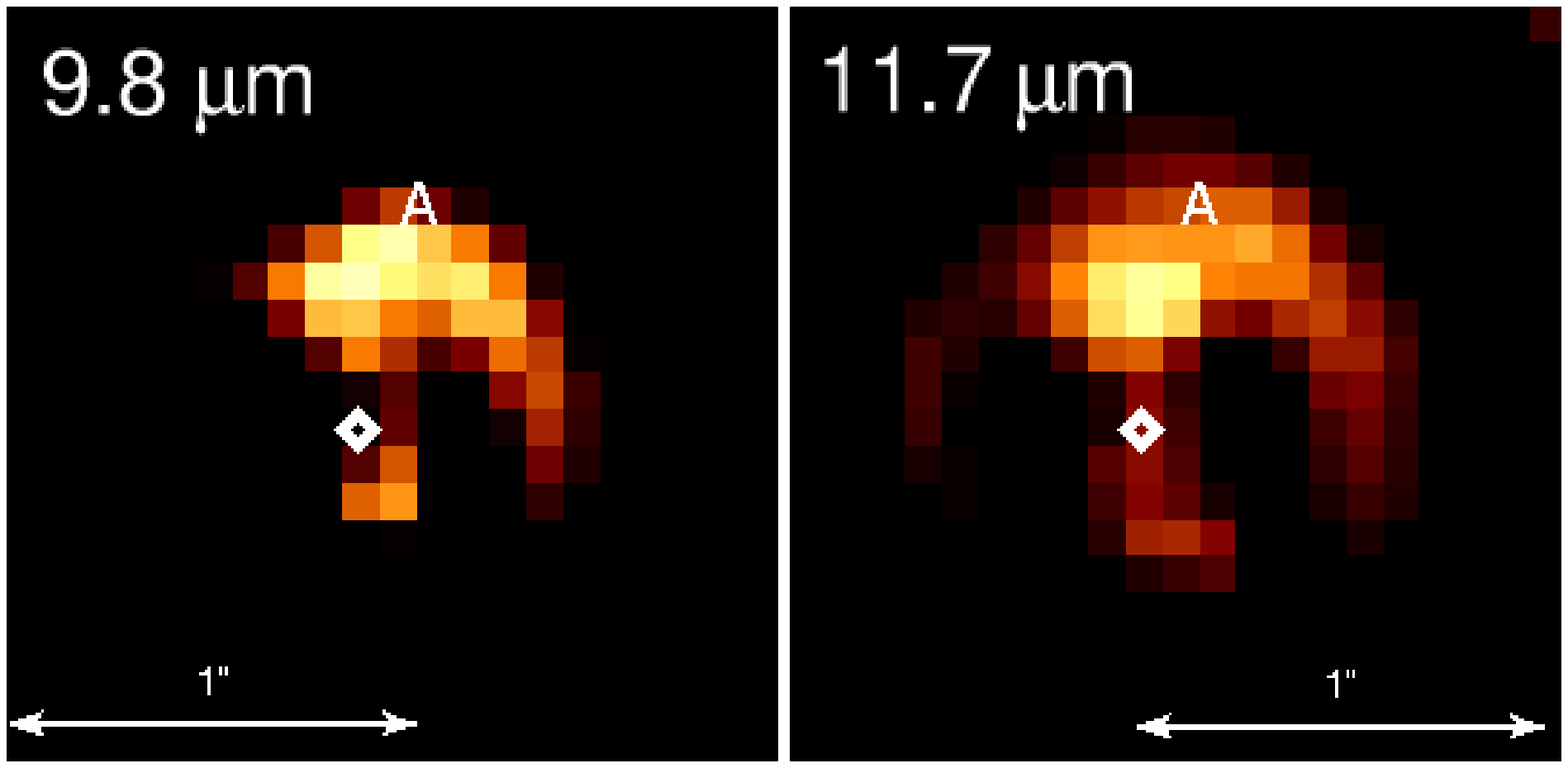} \\ 
\includegraphics[angle=0,width=4.0in]{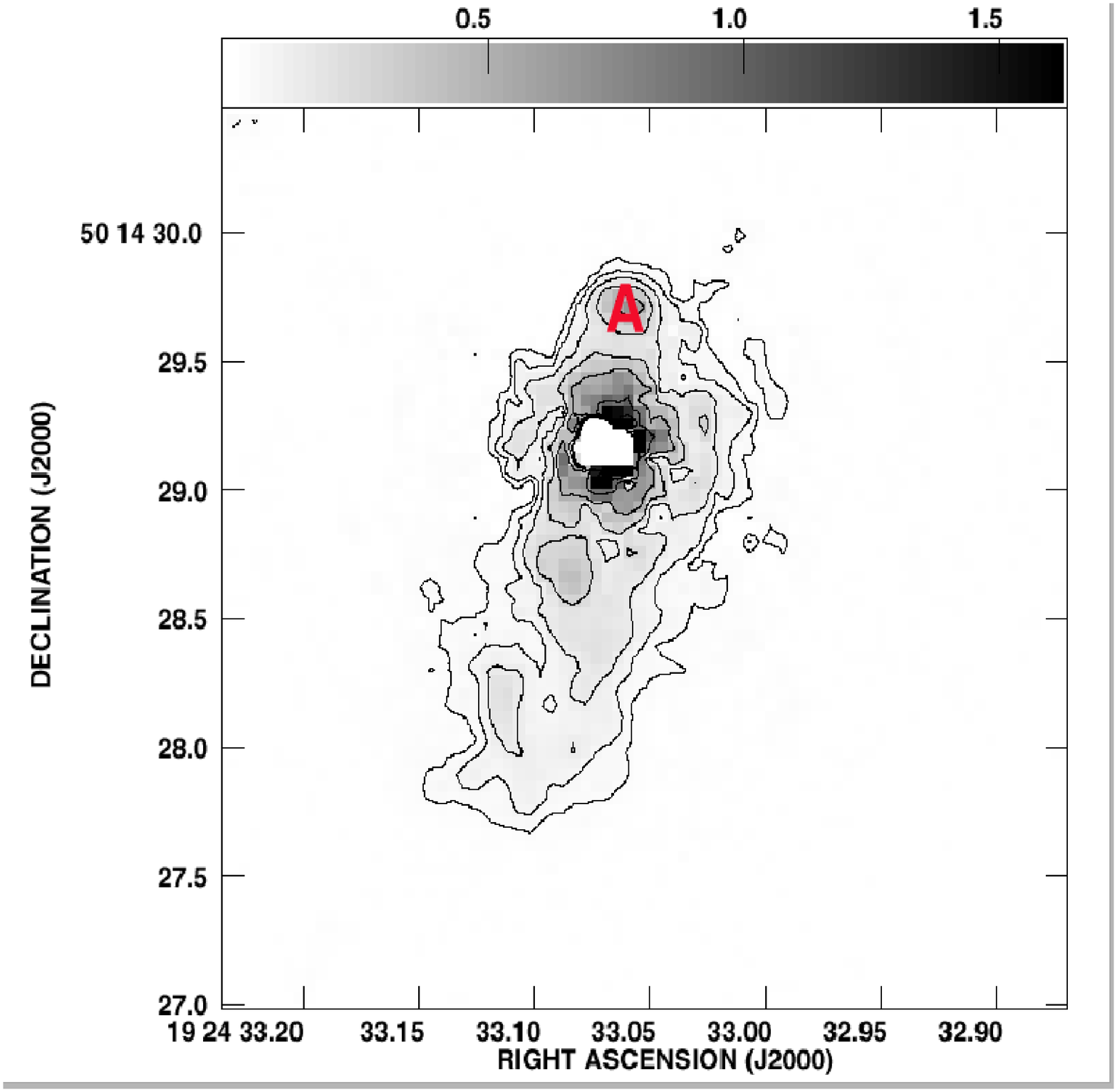} \\
\end{tabular}
\caption{Top: Faint residual structure ($\leq$2$\%$ of the peak) 
	left over after scaling and subtracting 
        the best-fit Gaussian model (convolved with the $\mu$ UMa PSF) 
        from the data. North is up and east is left in these images.  
        The diamond point represents the location of the stellar 
        components of the CH Cyg system in these images.  
        Bottom: An HST WFPC2 image of CH Cyg from August 1999 \citep{cro01} 
        is shown for comparison. 
        The residual extension in the mid-IR is roughly coincident
        with the northern knotlike/bipolar feature observed 
        by \citet{cro01} in the optical and the radio.} 
\label{fig:resids}
\end{figure}

We scaled and subtracted the best fit Gaussian model 
(convolved with the $\mu$ UMa PSF) from the data in 
order to reveal any asymmetric structure.  
Images of residual asymmetric structure are presented 
in Figure\,\ref{fig:resids}.  
The residual structure peaks in intensity   
$\simali$0.5$\arcsec$ north of the star (marked by a diamond) 
and is roughly coincident with the extension seen 
by \citet[]{cro01} at $\simali$0.7$\arcsec$ north 
of the star in HST WFPC2 images (marked by the letter A 
in Fig.~\ref{fig:resids}).  The structure also appears to 
be extended along a similar axis to the radio/optical jets.

\section{Modeling}
Since the residual asymmetric structure we observe around 
CH Cyg is faint compared to the symmetric Gaussian structure  
(peak fluxes only $\simali$2$\%$ that of the symmetric component) 
we have concentrated our modeling efforts 
on the symmetric component only.  
We model the IR emission of CH Cyg as an optically thin 
shell passively heated by the star.

We approximate the stellar photosphere by the Kurucz (1979) model 
spectrum for M6\,IIIe stars with an effective temperature of 
$\Teff$\,=\,3000$\K$. The dust is taken to be amorphous silicate 
since CH Cyg is a M star. Taking all grains to be spherical
with $a$ being their spherical radius, we assume a power-law 
dust size distribution $dn(a)/da \propto a^{-\alpha}$ 
which is characterized by 
a lower-cutoff $\amin$, upper-cutoff $\amax$ 
and power-law index $\alpha$.
For simplicity, we take $\amin$\,=\,0.01$\mum$ 
since smaller grains will undergo single-photon heating,
%(Draine \& Li 2001)
%while the observed IR spectral energy distribution
%of CH Cyg does not appear to show evidence for 
%stochastically heated dust, 
and $\amax$\,=\,1000$\mum$ since larger grains are 
not well constrained by the currently available 
IR photometry.

We approximate the dust spatial distribution 
with a modified power-law 
$dn/dr \propto \left(1-\rmin/r\right)^{\beta} 
\left(\rmin/r\right)^{\gamma}$
where $\rmin$ is the inner boundary of the envelope
which we take to be the location where silicate
dust sublimates. For CH Cyg with a luminosity
$\Lstar\approx 6900\,\Lsun$, 
submicron-sized silicate dust achieves 
an equilibrium temperature of $T\approx 1500\K$ 
and starts to sublimate at $r\approx 2\AU$. 
Therefore we take $\rmin=2\AU$. 
This functional form has the advantage that on one hand,
it behaves like a power-law $dn/dr \propto r^{-\gamma}$ at
larger distances ($r\gg\rmin$), and on the other hand
it peaks at $\rp = \rmin \left(\beta+\gamma\right)/\gamma$,
unlike the simple power-law which peaks at $\rmin$. 
The latter is unphysical since one should not expect 
dust to pile up at $\rmin$ where dust sublimates! 
We take $\gamma$\,=\,2 as expected from a stationary outflow
with a constant mass loss rate (e.g. see Habing, Tignon, $\&$ Tielens 1994).
But other $\gamma$ values have also been considered (see below), in view
of the time-variable nature of this object.
We take $\rmax$\,=\,200$\AU$ which is large enough for our
IR emission modeling purpose since there is very little dust
beyond this outer boundary which emit at the wavelengths
we are interested ($\lambda \simlt 200\mum$).
Therefore, we have only two free parameters: 
$\alpha$ -- the power-law exponent for the dust size
distribution and $\beta$ -- the dust spatial distribution
parameter which determines where the dust peaks.

Using the dielectric functions of ``astronomical silicates'' 
(Draine \& Lee 1984) and Mie theory (Bohren \& Huffman 1983), 
we calculate the absorption cross sections of amorphous 
silicate grains as a function of size, 
as well as their equilibrium temperatures
as a function of radial distance from the central star
(in thermal equilibrium with the illuminating starlight). 
We then obtain the dust model IR emission spectrum by 
integrating the dust emission over the entire size range 
and the entire envelope. 

\begin{figure}
\includegraphics[angle=0,width=\figwidth]{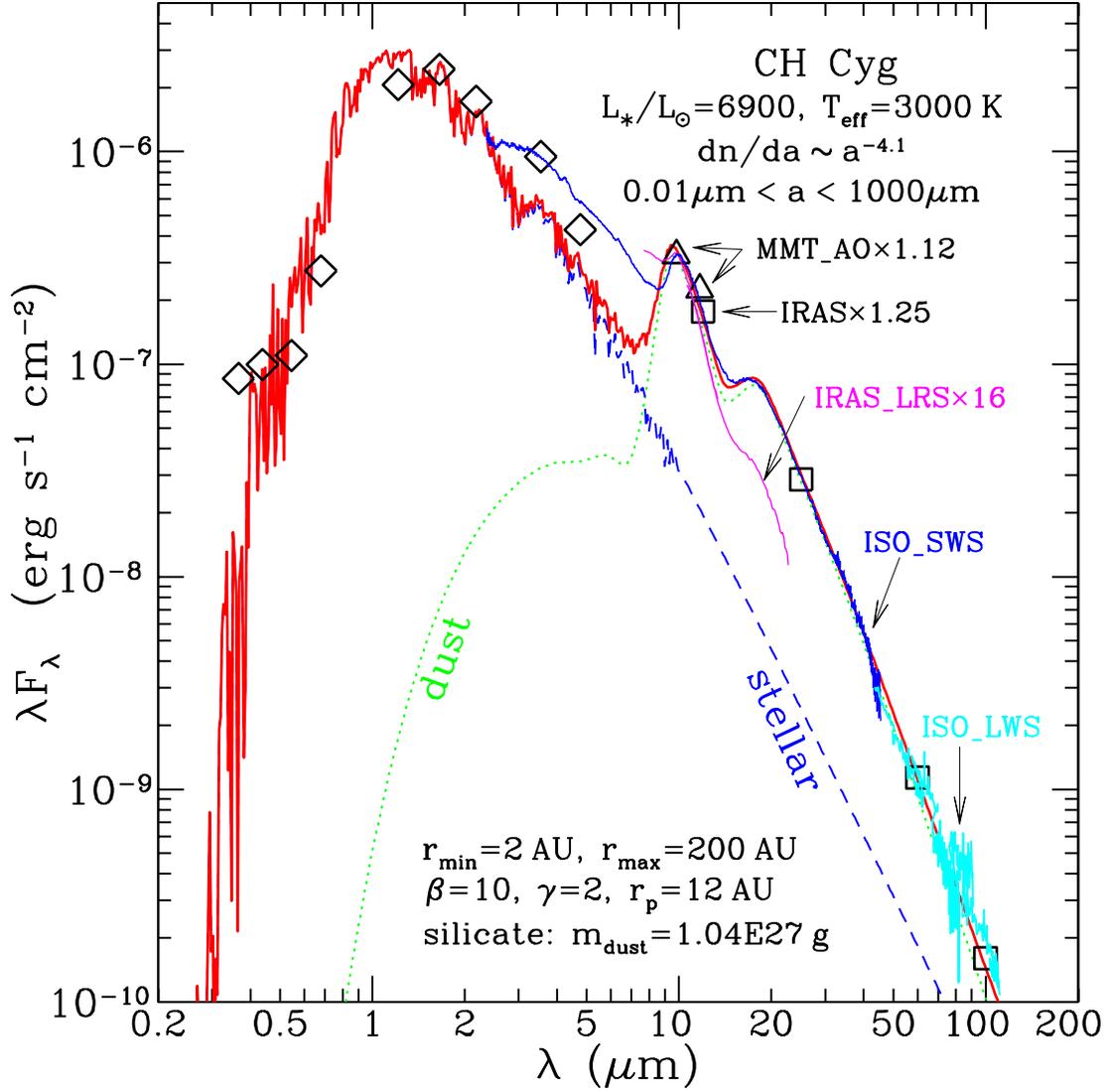}\\
\caption{IR emission and model fit to the CH Cyg dust shell. 
         Open black diamonds -- U, B, V, R, J, H, K, L,
         and M photometry from literature;
         open black triangles -- MMT-AO photometry 
         at 9.8$\mum$ and 11.7$\mum$
         (increased by a factor of 1.12); 
         open black squares -- IRAS photometry 
         at 12, 25, 60 and 100$\mum$ 
         (increased by a factor of 1.25).  
         Blue dashed line -- stellar photospheric spectrum;
         solid blue line -- ISO SWS spectrum;
         solid cyan line -- ISO LWS spectrum; 
         solid magenta line -- IRAS LRS spectrum
         (increased by a factor of 16).   
         The dust model is shown as a green dotted line, 
         while the solid red line plots the sum of the dust
         and the star.}
\label{fig:sed}
\end{figure}

The model spectrum is then compared with 
(1) the 9.8$\mum$ and 11.7$\mum$ 
photometry from our MMT AO images,
(2) the 12, 25, 60, and 100$\mum$ IRAS 
({\it Infrared Astronomical Satellite}) 
broadband photometry,
(3) the 7.7--22.7$\mum$ mid-IR low-resolution spectroscopy 
obtained with the IRAS {\it Low Resolution Spectrometer} 
(LRS; with a resolution of $\lambda/\Delta\lambda \approx 20$),
(4) the 2.5--45$\mum$ near- to mid-IR high-resolution 
spectroscopy obtained with 
the {\it Short Wavelength Spectrometer} 
(SWS; $\lambda/\Delta\lambda \approx 2000$)
on board the {\it Infrared Space Observatory} (ISO),
and (5) the 43--197$\mum$ mid- to far-IR medium-resolution
spectroscopy obtained with the ISO {\it Long Wavelength Spectrometer} 
(LWS; $\lambda/\Delta\lambda$\,$\approx$\,150--200).
Given that CH Cyg is a variable star, it is to our
great surprise that these observational data 
taken at different time epochs 
are in reasonably good agreement with each other: 
the MMT-AO and IRAS fluxes are only $\simali$12\%
and $\simali$25\%, respectively, smaller than the ISO SWS flux
(see Fig.\,\ref{fig:sed}).
The SWS spectrum is also in close agreement with the K, L, and M
photometry. But the IRAS LRS flux is lower than the SWS spectrum
by a factor of $\simali$16. 
We therefore take the ISO SWS and LWS spectra as
the comparison basis and increase the MMT-AO fluxes,
the IRAS broadband fluxes, and the IRAS LRS spectrum 
by a factor of 1.12, 1.25 and 16, respectively.
This brings the MMT-AO and IRAS data
into agreement with the ISO data.

The model which best fits the observed IR emission spectrum
and photometric points has $\alpha\approx$4.1 
and $\beta\approx$10 (see Fig.\,\ref{fig:sed}), 
with the dust distribution peaking at $\rp\approx 12\AU$, 
a total visual optical depth $\tau_V\approx 0.16$,
and a total dust mass $m_{\rm dust}\approx 1.04\times 10^{27}\g
\approx 5.2\times 10^{-7} M_{\odot}$. 
It is not easy to translate the total dust mass into
the mass loss rate which is actually considered as
one of the most important tasks in the study of stellar
evolution (e.g. see Willson 2000). To estimate the mass loss rate of CH Cyg,
we need to assume a gas-to-dust mass ratio $m_{\rm gas}/m_{\rm dust}$
and a gas outflow velocity $v$. With $m_{\rm gas}/m_{\rm dust}$\,=\,100
and $v$\,=\,$50\,{\rm km}\s^{-1}$ (based on the velocity of the Ca-II
absorption component which, according to Skopal et al. 1996, is
likely associated to the expanding circumstellar wind), we estimate 
the mass loss rate of CH Cyg to be 
$\approx 2.7\times 10^{-6}\,M_{\odot}\yr^{-1}$.

In order to compare the best fit model directly to our MMT AO images,
we convolved the model radial profile with our AO PSF image of $\mu$ 
UMa.  After convolution, the best fit model to the observed IR emission 
spectrum and photometric points 
produces an image appreciably larger than the MMT AO image.
We have considered models with a grid of parameters
($\beta$\,=\,2--20, $\gamma$\,=\,1--4, $r_{\rm max}$\,=\,25--250\,AU).
While multiple models fit the IR emission spectrum 
reasonably well, none of the models is able to simultaneously
replicate the AO image and the observed IR emission. 
For instance, the best fit model to the IR emission spectrum 
($\alpha\approx$4.1 and $\beta\approx$10) has a FWHM $\sim$ 0.4'' at 9.8 
$\mu$m (after convolution with the AO PSF), while the AO image of CH Cyg
has a FWHM $<$ 0.37''.  Given the extremely high Strehl ratios during these
observations ($\sim$100$\%$) and the resultant high image fidelity, 
this is a significant divergence!  These models are likely too simple to 
fully describe this complex system.  
The models assume only one star, with a steady state outflow.  
However, the CH Cyg system contains 3 stars -- we have not 
taken into account the disruptive effects upon the dust of 
the other two stellar components of this system.
The assumed functional form for the dust spatial
distribution may be too simplified.
We should also note that, as a variable star,
the spatial and size distributions of
the dust (which determine the surface brightness
distribution of the dust shell around CH Cyg) at 
the epoch when the AO images were taken can
be quite different from those when the IRAS
photometry and the ISO spectroscopy were obtained.
As a result, it is very unlikely that a dust model
will fit both the observed IR emission spectrum and the AO
images. Thus this is not a test of the validity of the model.

\section{Discussion}
Circumstellar dust in the CH Cyg system has previously 
been studied by Taranova and collaborators.  
With near- and mid-IR photometric observations, 
\citet{ts04} follow the properties of a circumstellar 
dust envelope around CH Cyg.  
According to \citet{ts04}'s model, 
mass was injected into the cool component's circumstellar
dust envelope around 1983 by an as yet unknown mechanism, 
coincident with the bipolar radio outflow observed by \citet{tay86}.
The optical depth of the dust envelope increased through 1992.  
By 1996, the dust envelope began to dissipate and had 
decreased to pre-mass injection optical depths by 2001.  

Taranova \& Shenavrin (2000, 2004) model the circumstellar 
dust envelope of CH Cyg as a single temperature shell 
around the cool component which varies in size 
and total dust mass during mass injection episodes.  
From 1983 to 2003, they claimed that the dust shell radius 
varied between $\simali$10$^{14}$--10$^{15}$\,cm, 
or from about 7--70\,AU
(but Bogdanov \& Taranova [2001] argue for a much large
extension: 15--15000\,AU).
We observe CH Cyg to be similarly extended, 
with a FWHM of $\simali$40--46\,AU.

%Using a more realistic model with dust density decreasing 
%as the square of increasing radius, \citet{bt01} modeled 
%the circumstellar material around the cool component of 
%CH Cyg (as observed in IRAS and ISO spectra) using the DUSTY code.  
%Their models have an inner radius of 15\,AU and 
%an outer radius of 15000\,AU.  
%Thus, the extension of CH Cyg we observe 
%is consistent with this model.  

The symmetric extended dust structure around CH Cyg is 
considerably larger than the orbits of the stellar components.  
In order to estimate the semimajor axis of the inner 756 day 
period binary, we adopt a white dwarf/hot component mass 
of $\simali$0.2\,$M_{\odot}$ and a red giant/cool component 
mass of $\simali$2\,$M_{\odot}$ \citep{hin93}.  
We calculate a semimajor axis estimate of $\simali$2\,AU. 
To estimate the semimajor axis of the outer 14.5 year period binary, 
we adopt an outer red giant mass of $\simali$2\,$M_{\odot}$ 
and find a semimajor axis estimate of $\simali$10\,AU.
The observed dust envelope extension of $\simali$40--46\,AU 
is larger than both the inner and outer binary orbits.
Thus, the dust envelope contains all three stellar 
components of this system.  The fact that all three stellar components
of this system lie within the dust envelope may be significant in explaining
why our AO dust shell images diverge 
from our spherically symmetric dust envelope models -- possibly, 
the stellar components of this system have a hand in shaping the 
dust envelope into something other than a classic, red giant wind 
spherical shell.  However, whatever specific contribution the stellar 
components have had on the shape of the dust envelope 
remains unresolved and unquantified.

After subtracting out the symmetric component 
of the emission, the residual emission appears 
extended along a similar north-south axis to 
those of the radio jets reported by \citet{cro01}.  
The residual structure peaks in flux $\sim$0.5$\arcsec$
north of the star and is roughly coincident to 
the knotlike structure observed by 
\citet[]{cro01} at 0.7$\arcsec$ north 
of the star in HST WFPC2 images 
(marked by the letter A in Fig.\,\ref{fig:resids}).  
The structure also appears to be extended along 
a similar axis to the radio/optical jets,
and is similar to the extent of the radio jets 
and ionized optical structure.  
However, while bipolar structure is observed in 
both the radio and optical, the mid-IR structure is unipolar.  
One possible explanation is that mid-IR emitting dust is 
denser towards the northern lobe of the bipolar outflow.
The outflow shocks and heats the dust towards the north,  
hence a bright mid-IR dust structure is observed.  Towards the south, 
the outflow is unimpeded by dust and, thus, has a larger 
optical and radio extent, but no mid-IR counterpart.  

\section{Conclusions}
The combination of adaptive optics with 
a deformable secondary produces very high 
Strehl images in the mid-IR and high PSF 
stability regardless of the seeing, airmass, 
or target brightness. Such PSF stability allows us to 
determine object properties with very high fidelity.  
Adopting the Hipparcos distance to CH Cyg of $\simali$270\,pc, 
we found a nearly spherically symmetric extension 
with a FWHM at 9.8$\mum$ of $\simali$40.5$\pm$2.7\,AU 
(0.15$\pm$0.01\,$\arcsec$) and a FWHM at 11.7$\mum$ 
of $\simali$45.9$\pm$2.7\,AU (0.17$\pm$0.01\,$\arcsec$).  
After subtracting out a model of the spherically 
symmetric emission, we found residual extension 
$\simali$0.5$\arcsec$ north of the star roughly coincident
with the optical knotlike feature observed by \citet[]{cro01}.

Adaptive optics at mid-IR wavelengths is a very promising
new technique that allows for uniquely stable PSFs 
and high Strehls. A high degree of PSF stability will 
eliminate morphological ambiguities due to poor 
(seeing-limited) PSF calibrations. 
Mid-IR AO should have a significant impact on 
any field where mid-IR imaging is possible.

\acknowledgements
These MMT observations were made possible with the hard
work of the entire Center for Astronomical Adaptive Optics (CAAO)
staff at the University of Arizona.  Graduate student Wilson Liu helped
run the MIRAC3 camera during the run. The adaptive secondary mirror is
a joint project of University of Arizona and the Italian National
Institute of Astrophysics - Arcetri Observatory. We would also like
thank the whole MMT staff for their excellent support and flexibility
during our commissioning run at the telescope.

The secondary mirror development could not have been possible without
the support of the Air Force Office of Scientific Research under grant
AFOSR F49620-00-1-0294.  BAB is supported by the NASA GSRP grant NNG04GN95H
and NASA Origins grant NNG056L71G.  LMC acknowledges support from NASA Origins
grant NNG056L71G and NSF CAREER award.
AL acknowledges support from the University of Missouri Summer 
Research Fellowship, the University of Missouri Research Board, 
and the NASA award P20436.
JHB acknowledges support from NSF grants AST-9987408 and AST-0307687.

\clearpage

\clearpage

\clearpage

\clearpage

\clearpage

\clearpage

\clearpage

\begin{thebibliography}{}
\bibitem[Biller et al.(2005)]{bil03}Biller, B. A., et al.\ 2005, \apj, 620, 450
\bibitem[Bogdanov \& Taranova(2001)]{bt01}Bogdanov, M.B., 
        \& Taranova, O.G. 2001, Astronomy Reports, 45, 797
\bibitem[Bohren \& Huffman(1983)]{bh83} Bohren, C.~F., \& 
         Huffman, D.~R.\ 1983, Absorption and Scattering of 
         Light by Small Particles, New York: Wiley
\bibitem[Close et al.(2003)]{close03}Close, L.M., et al.\ 2003, \apj, 598, L35
\bibitem[Corradi et al.(2001)]{cor01}Corradi, R.L.M., Munari, U., Livio, M., 
         Mampaso, A., Goncalves, D., \& Schwarz, H.E.  2001, \apj, 560, 912
\bibitem[Crocker et al.(2001)]{cro01}Crocker, M.M., et al.\ 2001, 
         MNRAS, 326, 781
\bibitem[Draine \& Lee(1984)]{dra84}Draine, B.T., \& Lee, H.M. 1984, 
         ApJ, 285, 89
%\bibitem[Draine \& Li(2001)]{dra01}Draine, B.T., \& Li, A. 
%         2001, ApJ, 551, 807
\bibitem[Eyres et al.(2002)]{eyr02}Eyres, S.P.S., et al.\ 2002, MNRAS, 335, 526
\bibitem[Galloway \& Sokoloski(2004)]{gs04}Galloway, D.K., 
        \& Sokoloski, J.L. 2004, \apj, 613, L61
\bibitem[Habing, Tignon, \& Tielens (1994)]{htt94} Habing, H.J., Tignon, J., \& Tielens, A.G.G.M.  1994, \aap, 286, 523
\bibitem[Hinkle et al.(1993)]{hin93} Hinkle, K.H., Fekel, F.C., Johnson, D.S.,
         \& Scharlach, W.W.G. 1993, \aj, 105, 1074
\bibitem[Hinz et al.(2000)]{hin00}Hinz P.M., et al.\ 2000, SPIE, 4006, 349
\bibitem[Hoffmann et al.(1998)]{hof98}Hoffmann, W. et al.\ 1998, 
        SPIE, 3354, 647
\bibitem[Karovska et al.(1998)]{kcm98}Karovska, M., Carilli, C.L., 
        \& Mattei, J.A. 1998, Journal of the American Association of 
Variable Star Observers, 26, 97
\bibitem[Kenyon(2000)]{ken00}Kenyon, S.J. 2000, in The Encyclopedia of 
        Astronomy and Astrophysics, ed. P. Murdin 
        (Bristol: Institute of Physics), 1630 
\bibitem[Kenyon(2001)]{ken01}Kenyon, S.J. 2001, in The Encyclopedia of 
        Astronomy and Astrophysics, ed. P. Murdin 
        (Bristol: Institute of Physics), 288 
\bibitem[Kurucz(1979)]{kur79}Kurucz, R.L. 1979, ApJS, 40, 1
\bibitem[Lucy(1974)]{lucy74}Lucy, L.B.  1974, \aj, 79, 745
\bibitem[Mikolajewski et al.(1987)]{mik87} Mikolajewski, M., Mikolajewska, J., 
\& Tomov, T.  1987, AP\&SS, 131, 733
\bibitem[Skopal et al.(1996)]{sko96}Skopal, A., et al.\
         1996, MNRAS, 327, 346
\bibitem[Sokoloski \& Kenyon(2003a)]{sk031}Sokoloski, J.L., \& Kenyon, S.J.  
        2003a, \apj, 584, 1021
\bibitem[Sokoloski \& Kenyon(2003b)]{sk032} Sokoloski, J.L., \& Kenyon, S.J.  
        2003b, \apj, 584, 1027
\bibitem[Taranova \& Shenavrin(2000)]{ts00}Taranova, O.G. \& Shenavrin, V.I. 
        2000, Astronomy Reports, 44, 460
\bibitem[Taranova \& Shenavrin(2004)]{ts04}Taranova, O.G. \& Shenavrin, V.I.
        2004, Astronomy Reports, 48, 813
\bibitem[Taylor et al.(1986)]{tay86}Taylor, A.R., Seaquist, E.R., 
        \& Mattei, J.A. 1986, Nature 319, 38
\bibitem[Wildi et al.(2003)]{wil03}Wildi, F.P., Brusa, G., 
        Lloyd-Hart, M., Close, L.M., \& Riccardi, A.  
        2003, SPIE, 5169, 17
\bibitem[Willson(2000)]{will00} Willson, L.~A.\ 2000, \araa,
	38, 573
\end{thebibliography}
\end{document}